\documentclass[prb,floats,aps,preprint,superscriptaddress,showpacs]{revtex4}
\usepackage{amsmath}
\usepackage{graphicx}

\newcounter{ss}
\newcommand{\Ssection}[2]{\subsection{\hskip -1.3em~\arabic{ss}\hskip 0.8em~#1}} 

\begin{document}
\date{\today}

\title{The inelastic hard dimer gas: a non-spherical model for granular matter.}

\author{Giulio Costantini}
\affiliation{Dipartimento di Fisica, Universit\`a di Camerino,
Via Madonna delle Carceri, I-62032 , Camerino, Italy and
INFM, Unit\`a di Camerino}
\author{Umberto Marini Bettolo Marconi}
\affiliation{Dipartimento di Fisica, Universit\`a di Camerino,
Via Madonna delle Carceri, I-62032 , Camerino, Italy and
INFM, Unit\`a di Camerino}
\author{Galina Kalibaeva}
\affiliation{Dipartimento di Fisica, Universit\`a La Sapienza,
P.le A. Moro 2, I-00185 Rome, Italy} 
\author{Giovanni Ciccotti}
\affiliation{Dipartimento di Fisica, Universit\`a La Sapienza and
INFM Unit\`a di Roma I, P.le A. Moro 2,~ I-00185 Rome, Italy}

\begin{abstract}
We study a two-dimensional gas of 
inelastic smooth hard dimers. Since the collisions between dimers
are dissipative, being characterized by a coefficient of restitution
$\alpha<1$, and no external driving force is present,
the energy of the system decreases in time and no stationary state
is achieved. However, the resulting non equilibrium state of the system
displays several interesting properties in close analogy with
systems of inelastic hard spheres, whose relaxational dynamics
has been thoroughly explored. 

We generalise to inelastic 
systems a recently method introduced
[G.Ciccotti and G.Kalibaeva, J. Stat. Phys. {\bf 115}, 701 (2004)] to
study the dynamics of rigid elastic bodies made up of
different spheres hold together by rigid bonds. 
Each dimer consists of two hard disks
of diameter $d$, whose centers are separated by a
fixed distance $a$. 
By describing the rigid bonds by means of holonomic constraints
and deriving the appropriate collision rules between dimers, we 
reduce the dynamics to a set of equations  
which can be solved by means of event driven simulation. 
After deriving the algorithm we study 
the decay of the total kinetic energy, and of the
ratio between the rotational and the translational kinetic energy
of inelastic dimers. We show numerically that 
the celebrated Haff's homogeneous cooling
law $t^{-2}$, describing how the kinetic energy of an inelastic
hard sphere system with constant coefficient of restitution
decreases in time, holds even in the case of these non spherical particles.
We fully characterize this homogeneous decay process in terms
of appropriate decay constants and confirm numerically
the scaling behavior of the velocity distributions.

\end{abstract}
\maketitle

\section{Introduction}
Over the past two decades much progress has been made in understanding
the dynamics of granular matter, i.e a collection of macroscopic particles
interacting via a short range repulsive potential in which energy is 
lost in inelastic 
collisions~\cite{gases,Jaeger,Kadanoff}. 
The industrial and technological importance
of products which are either powders or granulates is now becoming generally
recognized.  
Such a continued interest has unveiled 
a series of surprising phenomena and determined new challenges to physicists,
because of the simultaneous presence of many body effects and non thermal
fluctuations.
In fact, in granular materials such as sand, cereals, steel balls etc.
the ordinary temperature is irrelevant and consequently thermal equilibrium
is not realized. Due to inelastic collisions a granular system,
when not driven cools, i.e. the 
one particle velocity distribution tends to narrow 
in time~\cite{Goldhirsch,Haff}.
The inelastic hard sphere (IHS) model represents, perhaps, the simplest
description of a granular system and
most of the theoretical investigations are based on it.
A collection of 
IHS sufficiently rarefied is called granular gas, because its dynamics
consists of  ballistic motions  alternated with
instantaneous binary collisions which conserve total momentum.
The inelastic collisions are modeled by means of constant coefficient
of normal restitution $\alpha$, which determines the rate at which the kinetic
energy is dissipated. 
Therefore, we possess
nowadays a large amount of information concerning the IHS  
which  has become the standard reference model for granular matter
~\cite{Sela1,brey1,Deltour,Esipov,Noije1,Noije2,Sela2}.

A natural question to ask is whether the behaviors observed
in the IHS are generic or peculiar. 
The relevance of such an issue is not academic, since
many actual materials, such as rice, needles,
are far from being spherical but are elongated. As far as thermal systems are 
concerned the role of the shape of the constituent molecules 
in determining their thermodynamic behavior has been recognized 
since the pioneering work of Onsager \cite{Onsager}.
Entropic effects (packing) of sufficiently anisotropic molecules
(prolate ellipsoids, needles, spherocylinders, disks)
determine the formation of mesophases~\cite{Frenkel}.  
In granular matter, instead, the effect of the non spherical shape of the
particles is still largely unexplored in spite of its potential relevance
in many phenomena such as nematic ordering transitions or violations
of the energy equipartition. 
Few exceptions to such a trend are represented by the investigations
of refs. \cite{Zippelius,Poschel1,Luding,Villarruel}.

Since the computer simulation is a fundamental tool in the study
of granular systems, it is crucial to devise an efficient method 
to treat the collisions, which usually represents the most time consuming
part of a numerical code. The Event Driven (ED) method reduces such an effort
to the calculation of the collision times, since the interactions
in hard core systems occur only at the moment of collision.
Once the shortest collision time has been computed all the particle
positions are propagated inertially with the new velocities 
determined by the collision rules. The advantage over the fixed time step
method in the case of hard objects is clear since there is no 
algorithmic error in
the integration of the equations of motion.  However, in the case of 
non spherical bodies the calculation of the collision time is not
simple. Therefore, some authors resorted to a fixed time step method.
Recently, Ciccotti and Kalibaeva \cite{ciccotti} have shown that it is possible
to include a constraint, such as a fixed bond length between two 
elastic hard spheres, and preserve the advantages of the ED simulation.
Hereafter we illustrate how to extend the ED method to a model
of bond constrained inelastic hard disks.

%%%%%%%% LAY-OUT

The structure of the paper is the following. In section II we
define the model and derive the equations of motion for a 
pure system of inelastic dimers.  
We also generalise the method to mixtures of inelastic dimers and disks
and to systems of dimers in the presence of a fixed impenetrable wall.
In section III we present numerical results obtained by means of the ED
simulation. We discuss the properties of the homogeneous cooling state
of inelastic hard dimers and obtain numerically the relevant parameters
and the distribution functions.
In section IV we draw the conclusions and discuss the future
developments.

\section{Model and Algorithms}
\begin{figure}[htb]
{\includegraphics[clip=true,width=8.cm, keepaspectratio,angle=-90]
{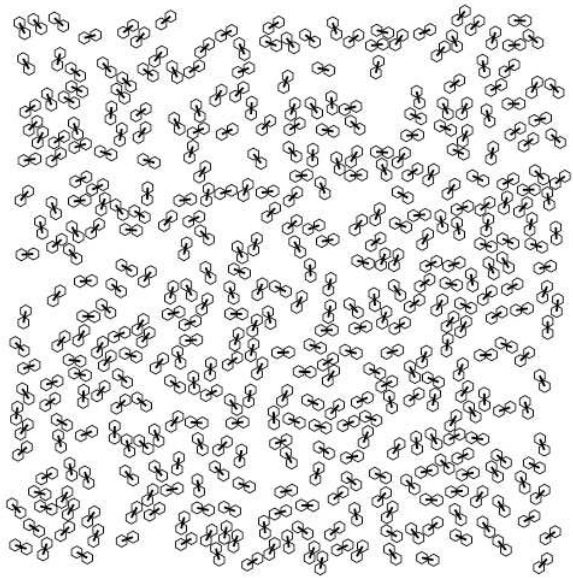}}
\caption
{Snapshot of a system of 400 dimers with $a=d$ and $\alpha=0.95$.}
\label{snap}
\end{figure}

In the present paper we discuss the properties of a system comprised
of $N$ dimers consisting of two identical hard disks of mass $m$ rigidly
connected. 
The diameter of the disks is $d$ and the distance between
their centers is $a$. In Fig.~\ref{snap} there is
 a snapshot of the system.
By describing  the rigid bond between the two
disks via an holonomic constraint~\cite{ciccotti} we are able to 
apply the ED numerical simulation technique.

Let us consider the two-dimensional motion of an assembly 
of rigid dimers. We make the assumption of rigid body dynamics:
\begin{itemize}
\item The duration of contact is negligible and the interaction
forces are high, so the velocity changes are nearly instantaneous
without notable change in positions.
\item The contact area is also negligible and the deformations are small
in the impact zone so that the impact occurs at a single point
of each body.
\item Double contacts cannot occur simultaneously.
\end{itemize}

The motion of each dimer, $A$, 
is described by the velocity of its center of mass
${\bf V}_A$ and by its angular velocity ${\bf \omega}_A$. The discontinuities
of linear and angular momentum occurring at a collision event are 
obtained by imposing the conservation of the total linear and angular
momentum and the law governing energy loss.

%%%%%%%%%%%%%%%%%%%%%%%%%%%%%%% FIG 1  Snapshot %%%%%%%%%%%%%%%%%%%%

%%%%%%%%%%%%%%%%%%%%%%%%%%%%%%% FIG 4  VELOCITY DISTRIBUTION %%%%%%%%%

\subsection{Free streaming}
In between collisions the dimers in the absence of external forces
perform an unperturbed roto-translational motion. As shown in 
ref.\cite{ciccotti} such a free streaming can be described by the
following parametric equations for the coordinates ${\bf r}_1(t)$
and ${\bf r}_2(t)$ of the centers of the two disks forming the dimer:

\begin{equation}
\begin{aligned}
{\bf r}_1(t)={\bf R}_A(0)+{\bf V}_A(0)t-\frac{1}{2}\Big[{\bf r}_2(0)-{\bf r}_1(0) \Big] \cos(\omega_A t)-
\frac{{\bf v}_2(0)-{\bf v}_1(0)}{2\omega_A}\sin(\omega_A t) \\
{\bf r}_2(t)={\bf R}_A(0)+{\bf V}_A(0)t+\frac{1}{2}\Big[{\bf r}_2(0)-{\bf r}_1(0) \Big] \cos(\omega_A t)+
\frac{{\bf v}_2(0)-{\bf v}_1(0)}{2\omega_A}\sin(\omega_A t) 
\end{aligned}
\label{streaming}
\end{equation}
where  ${\bf R}_A(0)$ e ${\bf V}_A(0)$ are respectively 
the position and the velocity of the center of mass of the
dimer A at the instant $t=0$.
The first two terms describe the translational motion, 
while the two last describe the rotational motion about the center of mass
of the dimer. The angular velocity $\omega_A$ is obtained by solving
the dynamics in the presence of the holonomic constraint
\begin{equation}
\chi=[{\bf r}_2(t)-{\bf r}_1(t)]^2-a^2=0
\label{holonomic}
\end{equation}
In fact differentiating with respect to $t$ eq.~(\ref{holonomic}) we obtain:
\begin{equation}
\begin{aligned}
\big[{\bf v}_2(t)-{\bf v}_1(t)]\cdot[{\bf r}_2(t)-{\bf r}_1(t)\big]
=-\omega_A\Big\{[{\bf r}_2(0)-{\bf r}_1(0)]\sin(\omega_A t) 
-[{\bf v}_2(0)-{\bf v}_1(0)]\frac{\cos(\omega_A t)}{\omega_A}\Big\}\\
\cdot \Big\{[{\bf r}_2(0)-{\bf r}_1(0)]\cos(\omega_A t)+
[{\bf v}_2(0)-{\bf v}_1(0)]\frac{\sin(\omega_A t)}{\omega_A}\Big\}=0
\end{aligned}
\label{omega2}
\end{equation}
Using eqs. (\ref{holonomic}) and (\ref{omega2}) we obtain
that angular velocity is constant and given by:
\begin{equation}
\omega_A=\sqrt{\frac{\big[{\bf v}_2(0)-{\bf v}_1(0)\big]^2}{a^2}}
\label{omega}
\end{equation}

\subsection{Collision times}
In order to give a self-contained account of the method we include
the determination of the collision times following the presentation
given in ref.\cite{ciccotti}.
The condition for the collision between two ``molecules'' A and B,
formed by the four atoms 1, 2 and 3, 4, respectively are:
\begin{equation}
\begin{aligned}
\big[{\bf r}_3(t)-{\bf r}_1(t)\big]^2=d^2\\
\big[{\bf r}_3(t)-{\bf r}_2(t)\big]^2=d^2\\
\big[{\bf r}_4(t)-{\bf r}_1(t)\big]^2=d^2\\
\big[{\bf r}_4(t)-{\bf r}_2(t)\big]^2=d^2\\
\end{aligned}
\label{distance}
\end{equation}
Substituting the equations (\ref{streaming})
for the position of each hard disk
as a function of time in the collision conditions (\ref{distance})
we get a set of equations in $t$, from which we take the smallest
one as the collision time of the two molecules.

\setcounter{subsection}{1}
\setcounter{ss}{1}
\Ssection{Dynamics of Collisions}

In each collision between two dimers four disks are involved, therefore 
we write the conservation laws for the total linear momentum
\begin{equation}
m({\bf v}_{1f}+{\bf v}_{2f}+{\bf v}_{3f}+{\bf v}_{4f})=
m({\bf v}_{1i}+{\bf v}_{2i}+{\bf v}_{3i}+{\bf v}_{4i})
\label{linear}
\end{equation}
and for the total angular momentum
\begin{equation} 
m({\bf r}_{1}\times{\bf v}_{1f}+{\bf r}_{2}\times {\bf v}_{2f}+
{\bf r}_{3}\times{\bf v}_{3f}+{\bf r}_{4}\times{\bf v}_{4f})=
m({\bf r}_{1}\times{\bf v}_{1i}+{\bf r}_{2}\times {\bf v}_{2i}+
{\bf r}_{3}\times{\bf v}_{3i}+{\bf r}_{4}\times{\bf v}_{4i})
\label{angular}
\end{equation}
where the subscripts 
$i$ and $f$ indicate pre-collisional
and post-collisional observable, respectively. Since 
the variables ${\bf r}_{j}$ ($j=1,4$) do not change during the instantaneous
collision, we do not need to distinguish between their pre-collisional
and post-collisional value.

Upon differentiating the constraint (\ref{holonomic}) we obtain relations 
between the velocities of the particles belonging to the same dimer:
\begin{equation}
\left \{\begin{aligned}
({\bf v}_{2i}-{\bf v}_{1i})\cdot{\bf r}_{12}=0 \\
({\bf v}_{2f}-{\bf v}_{1f})\cdot{\bf r}_{12}=0 \\
({\bf v}_{4i}-{\bf v}_{3i})\cdot{\bf r}_{34}=0 \\
({\bf v}_{4f}-{\bf v}_{3f})\cdot{\bf r}_{34}=0 \\ 
\end{aligned}
\right .
\label{constraints}
\end{equation}
with ${\bf r}_{jk}\equiv{\bf r}_{j}-{\bf r}_{k}$.
We simply satisfy eq. (\ref{linear}) by writing the change
of momenta of the two molecules A and B with the help 
of a vector ${\bf \Delta v}$, to be specified
in the following
\begin{equation}
\left \{\begin{aligned}
m ({\bf v}_{1f}+ {\bf v}_{2f})=m( {\bf v}_{1i}+{\bf v}_{2i})+
m {\bf \Delta v}\\
m({\bf v}_{3f}+ {\bf v}_{4f})=m( {\bf v}_{3i}+{\bf v}_{4i})-m{\bf\Delta v}\\
\end{aligned}
\right .
\label{totmomen2}
\end{equation}
In order to satisfy simultaneously  
eq. (\ref{totmomen2}) and
the constraints (\ref{constraints}) 
it has been shown, assuming that the colliding particles are $1$ and $3$, that one obtains the following expressions  \cite{ciccotti}
\begin{equation}
\left \{\begin{aligned}
\label{agg}
{\bf v}_{1f}={\bf v}_{1i}+{\bf \Delta v}-\frac{{\bf \Delta v}\cdot {\bf r}_{12}}{2a^2}{\bf r}_{12} \\
{\bf v}_{2f}={\bf v}_{2i}+\frac{{\bf \Delta v}\cdot {\bf r}_{12}}{2a^2}{\bf r}_{12} \quad \quad \quad \\
{\bf v}_{3f}={\bf v}_{3i}-{\bf \Delta v}+\frac{{\bf \Delta v}\cdot {\bf r}_{34}}{2a^2} {\bf r}_{34} \\
{\bf v}_{4f}={\bf v}_{4i}-\frac{{\bf \Delta v}\cdot {\bf r}_{34}}{2a^2} {\bf r}_{34} \quad \quad \quad
\end{aligned}
\right .
\end{equation}

The change of the angular momenta of the two colliding dimers are
\begin{equation}
\left \{\begin{aligned}
m ({\bf r}_{1}\times {\bf v}_{1f}+{\bf r}_{2} \times{\bf v}_{2f})=
m({\bf r}_{1}\times{\bf v}_{1i}+{\bf r}_{2}\times {\bf v}_{2i})
+m{\bf r}_{1}\times {\bf \Delta v}\\
m({\bf r}_{3}\times {\bf v}_{3f}+{\bf r}_{4} \times{\bf v}_{4f})=
m({\bf r}_{3}\times{\bf v}_{3i}+{\bf r}_{4}\times {\bf v}_{4i})
-m{\bf r}_{3}\times {\bf \Delta v}\\
\end{aligned}
\right .
\label{angmomen2}
\end{equation}
so that the  conservation of the total angular momentum at collision
implies that ${\bf \Delta v}$  is directed along the 
direction of the vector ${\bf r}_{13}$  connecting the centers of the two 
colliding disks, i.e
\begin{equation}
{\bf \Delta v}=\Delta v \frac{{\bf r}_{13}}{d}
\label{dv}
\end{equation}
Finally substituting eq. (\ref{dv}) into eq. (\ref{agg}) we arrive at
\begin{equation}
\label{agg2}
\left \{\begin{aligned}
{\bf v}_{1f}={\bf v}_{1i}+\Delta v \frac{{\bf r}_{13}}{d}-\Delta v\frac{{\bf r}_{13}\cdot {\bf r}_{12}}{2a^2d}{\bf r}_{12} \\
{\bf v}_{2f}={\bf v}_{2i}+\Delta v\frac{{\bf r}_{13}\cdot {\bf r}_{12}}{2a^2d}{\bf r}_{12} \quad \quad \quad \quad \\
{\bf v}_{3f}={\bf v}_{3i}-\Delta v \frac{{\bf r}_{13}}{d}+\Delta v\frac{{\bf r}_{13}\cdot {\bf r}_{34}}{2a^2d}{\bf r}_{34} \\
{\bf v}_{4f}={\bf v}_{4i}-\Delta v\frac{{\bf r}_{13}\cdot {\bf r}_{34}}{2a^2d}{\bf r}_{34}  \quad \quad \quad \quad
\end{aligned}
\right .
\end{equation}
In order to fix the amplitude $\Delta v$ we consider the change of total
kinetic energy occurring in a single collision event:
\begin{equation}
\begin{aligned}
\Delta E=
\frac{m}{2}({\bf v}^2_{1f}+{\bf v}^2_{2f}+{\bf v}^2_{3f}+{\bf v}^2_{4f})-
\frac{m}{2}({\bf v}^2_{1i}+{\bf v}^2_{2i}+{\bf v}^2_{3i}+{\bf v}^2_{4i})=\quad \quad \quad \quad \quad \\
m\Delta v^2\Big\{1-\frac{1}{4a^2d^2}\big[ ({\bf r}_{13}\cdot {\bf r}_{12})^2 + ({\bf r}_{13}\cdot {\bf r}_{34})^2\big]\Big\}-\frac{m\Delta v}{d}({\bf v}_{3i}-{\bf v}_{1i})\cdot {\bf r}_{13}
\end{aligned}
\label{de}
\end{equation}
For a perfectly elastic collision the condition $\Delta E=0$
determines $\Delta v$:

\begin{equation}
\Delta v=\frac{({\bf v}_{3i}-{\bf v}_{1i})\cdot {\bf r}_{13}/d}
{1-\frac{1}{4a^2d^2}\Big[ ({\bf r}_{13}\cdot {\bf r}_{12})^2 + ({\bf r}_{13}\cdot {\bf r}_{34})^2\Big]} \\
\label{elastic}
\end{equation}
To generalize the collision rule to the case of inelastic
collisions we consider the variation of the relative velocity of the two
colliding disks along the center to center direction:

\begin{equation}
\label{pip}
({\bf v}_{3f}-{\bf v}_{1f}) \cdot {\bf r}_{13}  =
({\bf v}_{3i}-{\bf v}_{1i}) \cdot {\bf r}_{13}- 2 \Delta v d \Big\{
1-\frac{1}{4a^2d^2}\big[ ({\bf r}_{13}\cdot {\bf r}_{12})^2 
+ ({\bf r}_{13}\cdot {\bf r}_{34})^2\big]\Big\} \\
\end{equation}
and impose the condition that the projection of the
relative velocity along ${\bf r}_{13}$  after the collision is
proportional to the projection of the relative velocity before the collision
via a coefficient of restitution $\alpha$
($0\leq \alpha\leq 1$),   
which takes into account the energy dissipation.
\begin{equation}
\label{pip1}
({\bf v}_{3f}-{\bf v}_{1f}) \cdot {\bf r}_{13}  =
-\alpha ({\bf v}_{3i}-{\bf v}_{1i}) \cdot {\bf r}_{13}
\end{equation}
In this case one sees that the rule (\ref{elastic}) is modified
\begin{equation}
\Delta v=\frac{1+\alpha}{2 d}
\frac{({\bf v}_{3i}-{\bf v}_{1i}) \cdot {\bf r}_{13}}
{1-\frac{1}{4a^2d^2}\big[ ({\bf r}_{13}\cdot {\bf r}_{12})^2 
+ ({\bf r}_{13}\cdot {\bf r}_{34})^2 \big]}
\label{inelastic}
\end{equation}

By substituting the relation (\ref{inelastic})  into  eq. (\ref{de}) we find
the corresponding energy loss in an inelastic collision
\begin{equation}
\Delta E=
-m\frac{1-\alpha^2}{4 d^2}
\frac{\big[({\bf v}_{3i}-{\bf v}_{1i})\cdot {\bf r}_{13}\big]^2}
{1-\frac{1}{4a^2d^2}\big[ ({\bf r}_{13}\cdot {\bf r}_{12})^2 + 
({\bf r}_{13}\cdot {\bf r}_{34})^2\big]} \\
\label{edissip}
\end{equation}
Let us compare the energy dissipated (\ref{edissip})
by two inelastic dimers with
that  associated with the collision of two inelastic hard disks of
diameter $d$, mass $m$ and coefficient of restitution $\alpha$:
\begin{equation}
\Delta E_{dd}=
-m\frac{1-\alpha^2}{4 d^2}
\big[({\bf v}_{3i}-{\bf v}_{1i})\cdot {\bf r}_{13}\big]^2 \\
\label{edissipdd}
\end{equation}
and notice that in the first case 
the amount dissipated depends on the relative orientation
of the two dimers, the largest dissipation occurring for  
head to head collisions,
where the two axes ${\bf r}_{12}$ and ${\bf r}_{34}$ are parallel to
${\bf r}_{13}$. This orientational dependence of  
the dissipation of dimers renders the model similar to a model 
of inelastic disks with a fluctuating coefficient of restitution.

\setcounter{subsection}{1}
\stepcounter{ss}
\Ssection{Collision between a dimer and a disk}
The behavior of mixtures of different grains
is also a subject of extreme interest. The dimer-disk mixture has not
been studied so far.  
It is a simple exercise to derive the collision rules in the case
of the collision between a dimer and a hard disk having the same
coefficient of restitution,
\begin{equation}
\left \{\begin{aligned}
\label{diskdimer}
{\bf v}_{1f}={\bf v}_{1i}+{\bf \Delta u}-\frac{{\bf \Delta u}\cdot {\bf r}_{12}}{2a^2}{\bf r}_{12} \\
{\bf v}_{2f}={\bf v}_{2i}+\frac{{\bf \Delta u}\cdot {\bf r}_{12}}{2a^2}{\bf r}_{12} \quad \quad \quad \\
{\bf V}_{df}={\bf V}_{di}-\frac{m}{M}{\bf \Delta u}\quad \quad\quad\quad\quad
\end{aligned}
\right .
\end{equation}
where we have assumed that the mass of the disk is $M$.
 ${\bf R}_d$ is the position of disk center and
$\sigma$ its diameter. We also introduced the  vector
$\Delta{\bf u}$:
\begin{equation}
{\bf \Delta u} =(1+\alpha_d)
\frac{({\bf V}_{di}-{\bf v}_{1i}) \cdot ({\bf R}_{d}- {\bf r}_{1})}
{1+\frac{m}{M}-\frac{1}{2a^2 C^2}
\big[({\bf R}_{d}-{\bf r}_{1})\cdot {\bf r}_{12}\big]^2}
\frac{({\bf R}_{d}- {\bf r}_{1})}{C^2}
\label{indiskdimer}
\end{equation}
with $C=(d+\sigma)/2$ and $\alpha_d$ is the coefficient of restitution. \\
In this case the total energy dissipation in a dimer-disk collision is
\begin{equation}
\Delta E_{d}=
-m\frac{1-\alpha_d^2}{2 C^2}
\frac{\big[({\bf V}_{di}-{\bf v}_{1i})\cdot
 ({\bf R}_{d}-{\bf r}_{1})\big]^2}
{1+\frac{m}{M}-\frac{1}{2a^2C^2}\big[ ({\bf R}_{d}-{\bf r}_{1})\cdot 
{\bf r}_{12} \big]^2} \\
\label{edissipdisk}
\end{equation}

\setcounter{subsection}{1}
\stepcounter{ss} 
\Ssection{Collision between a dimer and a smooth impenetrable wall}
A second extension of the previous methods concerns the collision of a dimer
against an impenetrable hard wall with a restitution coefficient
$\alpha_w$. Taking the limits $\sigma\to\infty$, $M\to\infty$ and ${\bf V}_{di}=0$, 
we find in this case:
\begin{equation}
\left \{\begin{aligned}
\label{walldimer}
{\bf v}_{1f}={\bf v}_{1i}
+{\bf \Delta w}-\frac{{\bf \Delta w}\cdot {\bf r}_{12}}{2a^2}{\bf r}_{12} \\
{\bf v}_{2f}={\bf v}_{2i}+\frac{{\bf \Delta w}\cdot {\bf r}_{12}}{2a^2}{\bf r}_{12} \quad \quad \quad \\
\end{aligned}
\right .
\end{equation}
with 
\begin{equation}
{\bf \Delta w} =-(1+\alpha_w)
\frac{{\bf v}_{1i} \cdot {\bf \hat{n}}}
{1-\frac{1}{2a^2}({\bf \hat{n}}\cdot {\bf r}_{12})^2}{\bf \hat{n}}
\label{indiskwall}
\end{equation}
and ${\bf \hat{n}}$ is a unit vector normal to the surface of the wall.\\
The relative total energy change is
\begin{equation}
\Delta E_{w}=
-m\frac{1-\alpha_w^2}{2}
\frac{({\bf v}_{1i}\cdot
 {\bf \hat{n}})^2}
{1-\frac{1}{2a^2} ({\bf \hat{n}}\cdot 
{\bf r}_{12} )^2} \\
\label{edissipwall}
\end{equation}

%%%%%%%%%%%%%%%%%%%%%%%%%%%%%%%%%%%%%%%%%%%%%%%%%%%%%%%%%%%%%

\section{Cooling Dynamics} 
In the present section we focus on the evolution of an initially 
homogeneous gas of inelastic ($\alpha<1$) hard dimers constrained
to move on a plane. The system consists of $N$
particles contained in a square  subject to periodic boundary conditions
and is not externally driven. Inelastic collisions between the dimers 
result in a loss of kinetic energy and thus determine a ``cooling''
of the gas. 

This problem has been attracting much interest 
since the inelastic collisions make the system to behave very different
from standard molecular fluids and plays a central role
in this sub-area of non equilibrium statistical mechanics.
The discovery of the homogeneous cooling state (HCS), characterized
by a one particle distribution function whose shape is self similar,
if the velocity is appropriately rescaled, has triggered a lot
of attention \cite{Gold2}.

 In order to put things in perspective
we take as a reference system,
a set of $N$ inelastic hard disks of diameter $\sigma$ in two 
spatial dimensions ($D=2$), since its properties have been
thoroughly studied. Such a system starting from
an equilibrium state of a corresponding elastic hard disk system,
cools down uniformly, and decreases its average
kinetic energy per particle 
$E_{kin}(t)=\frac{D}{2}~T(t)$ according to Haff's law~\cite{Haff}
\begin{equation}
T(t)=\frac{T_0}{(1+\gamma\nu_0 t)^2}
\label{haff}
\end{equation}
where $T(t)$ is the so called
granular temperature, $T_0$ the initial temperature,
$\nu_0$  is the equilibrium Enskog collision frequency
at $T_0$ and the non dimensional parameter $\gamma$ is: 
\begin{equation}
\gamma=\frac{(1-\alpha^2)}{2D}.
\label{gamma}
\end{equation}
Remarkably the homogeneous cooling law holds in any spatial dimension, $D$,
as far as smooth inelastic hard spheres are considered. 

HCS is characterized by uniform density
and velocity fields and is maintained only for a finite time, or for a finite
number of collisions, since it is linearly unstable to fluctuations.
After the initial homogeneous stage, the system enters an inhomogeneous
cooling state where a vortex structure develops in the velocity field
(shearing instability) followed by a density instability
(clustering)~\cite{Goldhirsch,brey2}.

A theoretical estimate of $\nu_0$ for elastic hard disks is provided
by the following Enskog expression
\begin{equation}
\nu_0=2 \sqrt{\pi} n \sigma\sqrt{\frac{T_0}{m}} g(\sigma)
\label{enskogfreq}
\end{equation}
where $n$ is the number of particles per unit area and
$g(\sigma)$ is the hard disk equilibrium pair correlation function evaluated at 
contact~\cite{Luding}:
\begin{equation}
g(\sigma)=\frac{1-\frac{7}{16}\eta}{(1-\eta)^2}
\end{equation}
and $\eta$ is the area fraction $\eta=\pi n \sigma^2/4$.
It is also useful, for future comparison, 
to express the number of atomic
collisions per particle, $\tau$, suffered by the particles until time $t$
by the phenomenological relation connecting to the inelasticity parameter $\gamma$ and the 
equilibrium collision frequency:

\begin{equation}
\tau(t)\equiv\frac{N_{atcoll}(t)}{N}=\frac{1}{\gamma}\ln(1+\gamma\nu_0 t)
\label{collisiontime}
\end{equation}

A second important property of the HCS concerns the velocity 
distribution function, which of course cannot be stationary 
due to the energy loss, but after a short transient assumes a 
form which depends on time only through the granular temperature,
$T(t)=\frac{1}{2}m v^2_0(t)$,

\begin{equation}
f({\bf v},t)=\frac{n}{v_0(t)^D} \Phi \Big(\frac{{\bf v}}{v_0(t)}\Big)
\label{pdv}
\end{equation}
with
\begin{equation}
v^2_0(t)=\frac{2}{D n}\int d^D v ~{\bf v}^2 f({\bf v},t)
\label{thermalvelocity}
\end{equation}

The scaling function, $\Phi(z)$ is time independent in the HCS and in the 
limit of small dissipation approaches a Gaussian, i.e. 
$\Phi(z)\simeq \pi^{-D/2}\exp(-z^2)$. For smaller values of
$\alpha$ a perturbation theory about the elastic state accounts
for the departure from the Maxwellian~\cite{Noije2}.

Is the scenario described above preserved in a gas of hard
inelastic dimers?  In other words can we find 
constant values $\gamma^{eff}$ and $\nu_0^{eff}$ such that 
eqs.~(\ref{haff}) and  (\ref{collisiontime}) still hold
with these effective parameters? Secondly, do the distribution functions
of the translational and rotational components have the scaling 
property~(\ref{pdv})?
If the answer is affirmative the inelastic hard dimer system possesses
a homogeneous cooling state, whose importance for the study
of granular gases has been stressed by various
authors ~\cite{Gold2,brey2,brito2,Nakanishi,Poschel,Puglio}.

\begin{figure}[htb]
{\includegraphics[clip=true,width=10.cm, keepaspectratio,angle=0]
{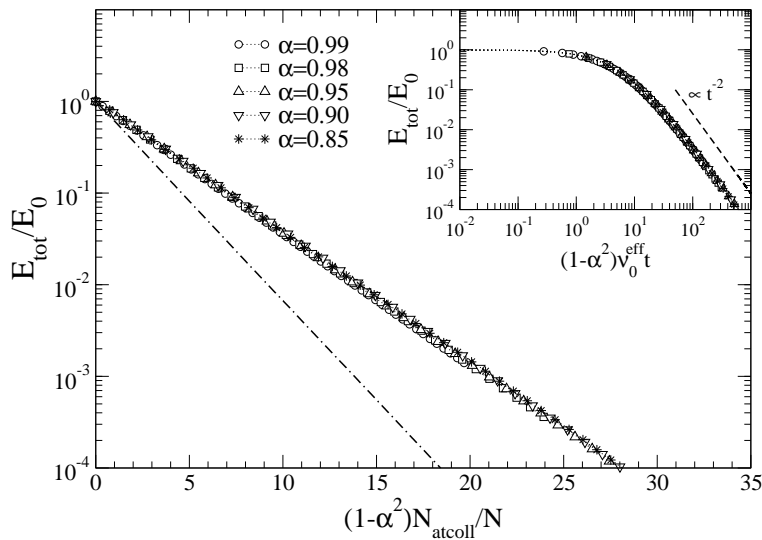}}
\caption
{Data collapse  for  dimers having different  
coefficients of restitution obtained by plotting the
decay of total energy as a function of $(1-\alpha^2)N_{atcoll}/N$, i.e. the total 
number of collisions  suffered by a molecule
multiplied by the
inelasticity factor $(1-\alpha^2)$. 
The dashed line corresponds to the energy decay of the
reference hard disk system and shows a faster decay.
In the inset 
the same quantities are shown as functions of the rescaled time $\nu_0^{eff}t$, showing the
expected Haff's inverse power law decay.}
\label{energytot}
\end{figure}

The evolution of the inelastic dimer system
can be characterized in terms of the
average values per particle of the total kinetic energy,  
the translational and the rotational kinetic energy. 
The last two quantities averaged over the particles define the 
partial granular temperatures:
\begin{equation}
T_{tr}(t)=<E_{tr}(t)>=\frac{m}{N}\sum_{A=1}^{N} \big[{\bf V}_A(t)-
\langle{\bf V}_A(t)\rangle \big ]^2
\end{equation}
\begin{equation}
T_{rot}(t)=2<E_{rot}(t)>=\frac{I}{N} \sum_{A=1}^{N}{\omega_A^2(t)}
\end{equation}
where ${\bf V}_A(t)$ is the center of mass velocity of a single dimer $A$,
 $\omega_A$ its angular velocity and $I=m a^2/2$ its moment of inertia.  
Our results for the
decay of the average total kinetic energy 
$E(t)=T_{tr}(t)+\frac{1}{2 }T_{rot}(t)$ are shown Fig.~\ref{energytot}.
We observe that the cooling process occurs according to the same universal 
inverse power law $t^{-2}$ which characterizes the inelastic hard 
sphere systems. In addition, when plotted as a function of the total
number of atomic collisions suffered on the average by a dimer times the inelasticity
parameter $(1-\alpha^2)$, the curves corresponding to different
values of $\alpha$ nicely fall one onto the other. However, the 
slope characterizing the various dimer systems is different,
being in fact slower, from the
corresponding slope of the hard disk systems. This feature represents the
fingerprint of the structure of the dimers.

\begin{figure}[htb]
{\includegraphics[clip=true,width=10.cm, keepaspectratio,angle=0]
{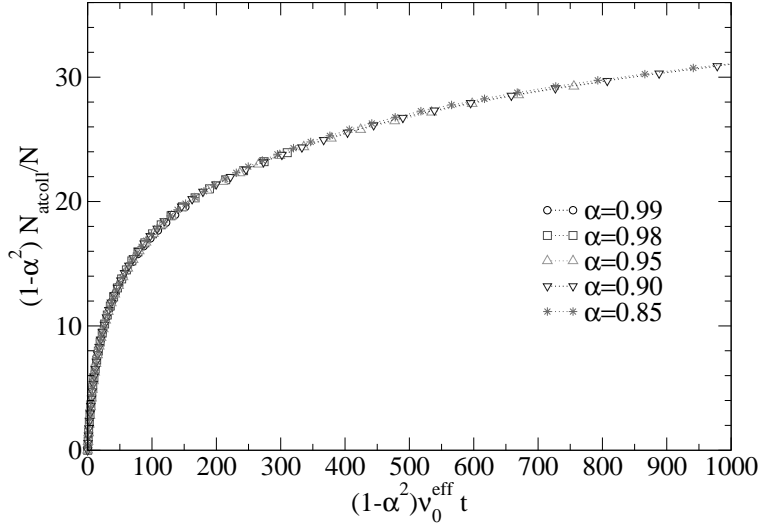}}
\caption
{Average number of atomic collisions per dimer rescaled
by the inelasticity factor $(1-\alpha^2)$ as a function of the rescaled time $\nu_0^{eff}t$ for various
values of the inelasticity and for a center-to-center distance $a=d$.}
\label{ncollisions}
\end{figure}

A second remarkable feature is the validity of the logarithmic 
law~(\ref{collisiontime}) (see Fig. \ref{ncollisions}), with the 
effective parameters $\nu_0^{eff}$ and $\gamma^{eff}=(1-\alpha^2)/(2 D_e)$
shown in table I. Notice that
$D_e=D$ in the case of inelastic 
$D$-dimensional smooth hard spheres.
 We note that while $D_e$ is nearly independent
of the elongation, $a$, of the dimer,  the initial collision frequency
$\nu_0^{eff}$ does depend on $a$, being connected with the total cross section
of this object.

%%%%%%%%%%%%%% FIGURA ENERGIA TOTALE %%%%%%%%%%%%%%%%%%%%%%%%%%%%%%%
%% CASO distanza centri=1 sigma

%%%%%%%%%%%%%%
%%%%%%%%%%%%%% Collision time %%%%%%%%%%%%%%%%%%%%%%%%%%%%%%%
%% CASO distanza centri=1 sigma

The entries of table I have been obtained  by fitting the numerical
data with the logarithmic law  
eq.~(\ref{collisiontime}) with $D_e$ and $\nu_0^{eff}$ used as fitting 
parameters. Our data analysis reveals that in all cases considered the fit
is very good and that $D_e$ is very close to $3$.

%\subsection{Heuristic derivation of cooling rate for dimers}

We use a simple heuristic argument in order to derive the slower cooling rate
of the hard dimers with respect to the disks. 
Let us compare formulae (\ref{edissip}) and (\ref{edissipdd}), representing
the energy dissipated in a collision between two dimers 
and two disks, respectively. 
What renders the behavior of the two systems different,
when the coefficient of restitution $\alpha$ of both is the same,
is the presence of the angular weighting factor, $f$:
\begin{equation}
f=\frac{1}{2-\frac{1}{2a^2d^2}\big[ ({\bf r}_{13}\cdot {\bf r}_{12})^2 + 
({\bf r}_{13}\cdot {\bf r}_{34})^2\big]} \\
\label{angularf}
\end{equation}
We can rewrite eq.~(\ref{edissip}) as
\begin{equation}
\Delta E=
-m\frac{1-\alpha^2}{2 d^2} f
\big[({\bf v}_{3i}-{\bf v}_{1i})\cdot {\bf r}_{13}\big]^2
\label{edi1}
\end{equation}
We consider the average value of eq.~(\ref{edi1}) over many collisions,
here indicated with the symbol $<\cdot>$ and assume the following 
factorization
\begin{equation}
\langle\Delta E \rangle=
-m\frac{1-\alpha^2}{2} \langle f \rangle \frac{1}{d^2}\langle \big[({\bf v}_{3i}-{\bf v}_{1i})\cdot {\bf r}_{13}\big ]^2 \rangle
\label{edi2}
\end{equation}
The last average turns out to be proportional to the average total
energy, i.e.:
  
\begin{equation}
\frac{1}{d^2}
\langle\big[({\bf v}_{3i}-{\bf v}_{1i})\cdot {\bf r}_{13}\big]^2 \rangle \simeq \frac{ \langle E \rangle}{m}
\label{edi3}
\end{equation}
which gives
\begin{equation}
\langle\Delta E \rangle=
-\frac{1-\alpha^2}{2} \langle f \rangle  \langle E \rangle.
\label{edi4}
\end{equation}

\begin{figure}[htb]
{\includegraphics[clip=true,width=10.cm, keepaspectratio,angle=0]
{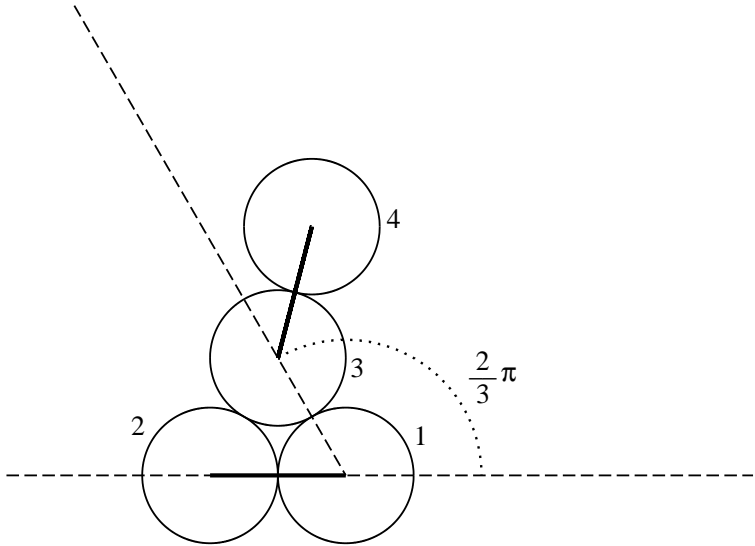}}
\caption
{Allowed angle between the directions ${\bf r}_{13}$ and  ${\bf r}_{12}$  
when two dimers are at contact.
Since the dimers cannot overlap
only the angles  between $-2\pi/3$ and $2\pi/3$ are allowed.}
\label{geometry}
\end{figure}

\begin{figure}[htb]
{\includegraphics[clip=true,width=8.cm, keepaspectratio]
{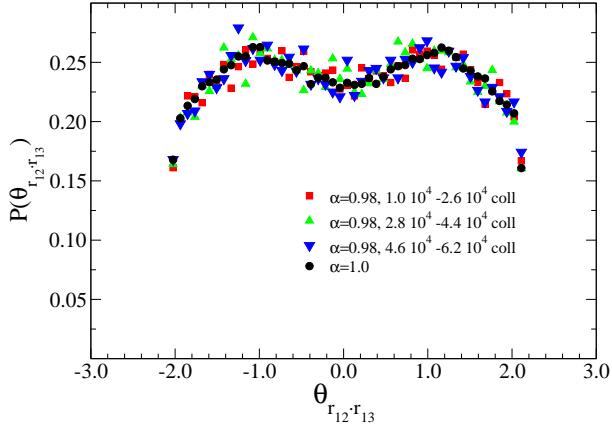}}
\caption
{Distribution of the angle between ${\bf r}_{12}$ and ${\bf r}_{13}$
measured in subsequent intervals, each containing   
$10^4$ collisions. The distribution
has been obtained for 900 dimers,
$\alpha=0.98$. The elastic case $\alpha=1.0$ is also shown
(bullets) for comparison.}
\label{angq}
\end{figure}

We compute $\langle f\rangle$, assuming
a dimer elongation $a=d$. It is straightforward to see that
the angles formed by the directions 
${\bf r}_{13}$ and  ${\bf r}_{12}$ or ${\bf r}_{34}$ are subjected to
the constraint, required by the condition of non overlapping two dimers,
that only the angles comprised between 
$-2\pi/3$ and $2\pi/3$ are allowed (see Fig. \ref{geometry}). 
We further assume a uniform distribution of these angles 
in the above interval,
a guess which is confirmed numerically. As shown in Fig. \ref{angq},
the computed distribution is almost uniform.
Using this fact, we
perform the average of formula~(\ref{angularf}) 
obtaining the value  $\langle f\rangle=0.64$. With this value we obtain
the effective ``dimension''  $D_e=D/\langle f \rangle=3.12$ 
which is compatible with the numerical values shown in table I.
Intuitively this result is consistent with the idea that the energy
stored in the rotational degree of freedom is only indirectly affected 
in the dissipation process and provides a kind of reservoir slowly dissipated. 
%%%%%%%%%%%%%% FIGURA Ratio energie  %%%%%%%%%%%%%%%%%%%%%%%%%%%%%%%

%%%%%%%%%%%%%%%%%%%%%%%%%%%%%%% FIG 4  NON EQUIPARTITION %%%%%%%%%%%%%%%%%%%%

%%%%%%%%%%%%%%%%%%%%%%%%%%%%%%% FIG temp1 %%% Q %%%%%%%%%%%%%%%%

\subsection{Energy equipartition}

One of the most peculiar features of granular systems is the
lack of energy equipartition among different kinetic
degrees of freedom. Such a phenomenon has been observed both in driven
and undriven systems of smooth and rough hard-sphere mixtures
\cite{Daniela,Dufty,Pagnani,Ciro} and during the 
homogeneous cooling of a gas of inelastic needles \cite{Zippelius}.

In the case of dimers, the translational 
and rotational degrees of freedom
satisfy the equipartition of energy only when $\alpha=1$, as required by 
equilibrium statistical mechanics.  
We have evaluated numerically this property by fitting the time series
for the ratio $K(\tau)=E_{tr}/E_{rot}$ by means of the linear least-square fit
$\bar{K}(\tau)=a_0+a_1 \tau$. We have performed 
a t-Student statistical test of the data to determine
the $99\%$ confidence intervals for $a_0$ and $a_1$~\cite{confidence}.  
The result of such a test shows that $a_1$ is consistent
with the value zero (see table II).
As already established for mixture of inelastic hard spheres~\cite{Dufty}
the ratio $K$ deviates from the corresponding elastic value 
($K=2$), but attains a constant ratio during the homogeneous cooling regime.
We observe  from table II that the larger the 
inelasticity the larger the breakdown of energy equipartition, and the 
longer the dimer the larger the deviation.

\subsection{Velocity distribution functions}

\begin{figure}[htb]
{\includegraphics[clip=true,width=6.cm, keepaspectratio,angle=0]
{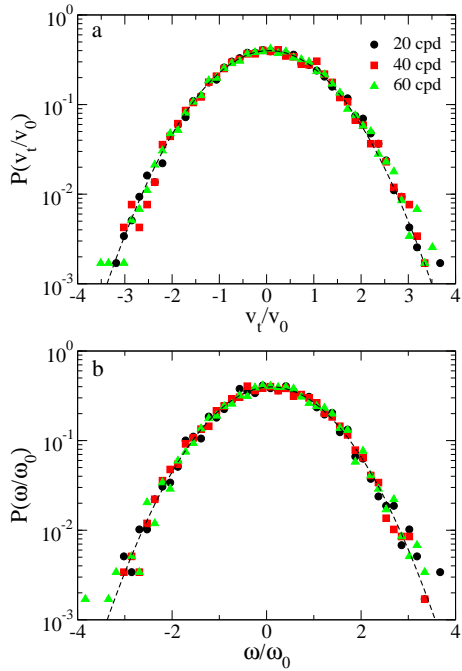}}
%{\includegraphics[clip=true,width=8.cm, keepaspectratio,angle=0]
%{dsvel-rot-3600-0_99-dil-1_0.eps}}
\caption{a) Normalized translational velocity distribution function
for dimers of elongation $a=d$ after different 
number of collisions per dimer (cpd).
b) Normalized rotational velocity distribution functions for the same
case. A gaussian curve (dashed line) is plotted for comparison.}
\label{distribtrasl}
\end{figure}

We turn now attention to the velocity distribution functions. As discussed
above one of the characterizing features of the HCS is the existence
of a time independent scaling function $\Phi(z)$. We have obtained 
numerically the related velocity distribution functions
for the two translational components of the center of mass
velocity and for the rotational velocity. These (normalized) distributions
are displayed in Figs.~\ref{distribtrasl}
and compared with the Gaussian. We have not tried
to measure departures from the Maxwellian, a task
beyond the scope of the present work. Moreover, it is known from the
literature~\cite{Goldshtein,Noije2} 
that in the case of inelastic hard-spheres these
departures are small in the range $0.6<\alpha<1$ so
that the Gaussian remains a valid
approximation in the HCS of inelastic hard spheres. We guess 
from the present data that such a behavior remains true even in the case
of the dimers.

To summarize there is evidence
that inelastic hard dimers display a normal behavior in the HCS 
even with respect to the nature of their velocity distribution functions
which are nearly Gaussian. Thus the analogy with the hard-sphere system 
is complete.

%%%%%%%%%%%%%% PDV %%%%%%%%%%%%%%%%%%%%%%%%%%%%%%%
%% CASO distanza centri=1 d

%%%%%%%%%%%%%%%%%%%%%%%%%%%%%%%%%%%%%%%%%

\section{Conclusion and Outlook}

We have investigated the dynamics of an undriven gas of inelastic 
hard dimers.
The first achievement of this work has been 
to formulate properly the dynamics of pure inelastic dimer systems
and that of 
a binary mixture of inelastic dimers and disks. 
 Using such results we treated a pure gas of dimers with the ED algorithm.
To validate the model and uncover some new physics 
we have then studied numerically the relaxation
dynamics of such a system during the homogeneous cooling stage
for various choices of the physical parameters.
By considering the total,
translational and rotational kinetic energies
of the dimers, we have found that 
their decay,
following an initial short transient, agrees, 
as far as the system remains homogeneous, with the one observed
in the case of a gas of IHS, that is Haff's cooling law.
The different geometrical shape manifests itself in determining
an identical pattern of cooling, although with a slower cooling rate
which is related with the internal structure of the dimers.
Moreover, we have found that the energy
equipartition principle does not hold, although the ratio
between
rotational and translational energy is constant and almost equal 
to the equipartition value. 

The one-particle velocity distribution functions of the system 
remain always close to the Maxwellian with properly scaled parameters 
both in the case of the center of mass translational 
and rotational degrees of freedom.
These observations allow us to conclude that a HCS exists also
in the case of inelastic hard dimers.

Preliminary work 
has also indicated interesting trends in the stages following the homogeneous
cooling regime, in particular we have observed that 
on a longer time scale the system breaks its homogeneity showing
vortex structures in the velocity field (shearing instability) and
clustering. Both phenomena have a counterpart in systems
of spherically symmetric particles. 
However, a careful analysis of these results requires further work.

\section{Acknowledgments}
U.M.B.M. acknowledges the support of the 
Project Complex Systems and Many-Body Problems
Cofin-MIUR 2003 prot. 2003020230.

\newpage
\begin{table}
\caption{\label{tab:table1} Dependence of the parameters $D_e$ and $\nu_0^{eff}/\nu_0$ on the  coefficient of restitution and on the elongation 
of the dimer. For comparison the hard disk theoretical values are $D_e=2$ and
as a numerical check we applied the same procedure as in the case of dimers and obtained $D_e=1.998$ at area fraction $\pi n \sigma^2/4=5.7\times10^{-4}$.}
\begin{ruledtabular}
\begin{tabular}{cccc}
\multicolumn{1}{c}{\rule[-0.3cm]{0mm}{0.3cm}}&
\multicolumn{1}{c}{$\alpha$}&
\multicolumn{1}{c}{$D_{e}$}&
\multicolumn{1}{c}{$\nu_0^{eff}/\nu_0$}\\

\hspace{1.0cm} & $0.99$ & $3.003$ & $1.07$\\
\hspace{1.0cm}& $0.98$ & $3.013$ & $1.08$\\
$a/d$ = 0.5 & $0.95$ & $3.028$ & $1.08$\\
\hspace{1.0cm} & $0.90$ & $3.079$ & $1.11$\\

\multicolumn{1}{c}{\rule[-0.3cm]{0mm}{0.4cm}}&
\multicolumn{1}{c}{$0.85$}&
\multicolumn{1}{c}{$3.109$}&
\multicolumn{1}{c}{$1.13$}\\ 
 
\hline

\multicolumn{1}{c}{\rule[0.1cm]{0mm}{0.5cm}}&
\multicolumn{1}{c}{$0.99$}&
\multicolumn{1}{c}{$2.987$}&
\multicolumn{1}{c}{$1.26$}\\ 

\hspace{1.0cm} & $0.98$ & $3.014$ & $1.25$\\
$a/d$ = 1.0  & $0.95$& $3.038$ & $1.28$\\
\hspace{1.0cm} & $0.90$ & $3.086$ & $1.32$\\

\multicolumn{1}{c}{\rule[-0.3cm]{0mm}{0.4cm}}&
\multicolumn{1}{c}{$0.85$}&
\multicolumn{1}{c}{$3.041$}&
\multicolumn{1}{c}{$1.31$}\\ 

\hline

\multicolumn{1}{c}{\rule[0.1cm]{0mm}{0.5cm}}&
\multicolumn{1}{c}{$0.99$}&
\multicolumn{1}{c}{$3.004$}&
\multicolumn{1}{c}{$1.45$}\\ 

\hspace{1.0cm} & $0.98$ & $3.031$ & $1.47$\\
$a/d$ = 1.5  & $0.95$ & $3.059$ & $1.50$\\
\hspace{1.0cm} & $0.90$ & $3.050$ & $1.48$\\

\multicolumn{1}{c}{\rule[-0.1cm]{0mm}{0.35cm}}&
\multicolumn{1}{c}{$0.85$}&
\multicolumn{1}{c}{$3.102$}&
\multicolumn{1}{c}{$1.50$}\\

\end{tabular}	
\end{ruledtabular}
\end{table}

\newpage

\begin{table}
\caption{\label{tab:table2} Dependence of the ratio $K$
between translational and rotational energy, 
on coefficient of restitution and elongation. 
%The $99\%$ confidence intervals $\Delta a_0$ and $\Delta a_1$ are also shown.}
The values $\Delta a_0$ and $\Delta a_1$, concerning the $99\%$ confidence intervals $a_0\pm\Delta a_0$ and $a_1\pm\Delta a_1$, are also shown.}
\begin{ruledtabular}
\begin{tabular}{cccccc}
\multicolumn{1}{c}{\rule[-0.3cm]{0mm}{0.3cm}}&
\multicolumn{1}{c}{$\alpha$}&
\multicolumn{1}{c}{$a_0$}&
\multicolumn{1}{c}{$\Delta a_0$}&
\multicolumn{1}{c}{$a_1$}&
\multicolumn{1}{c}{$\Delta a_1$}\\

\hspace{1.0cm} & $1.0$ & $1.999$ & $1\times 10^{-3}$ & $-2.86\times 10^{-6}$ &
$1.2\times 10^{-6}$\\
$a/d$ = 0.5 & $0.98$ & $2.014$ & $1\times 10^{-3}$ & $-2.82\times 10^{-5}$ &
$3.5\times 10^{-6}$\\
\hspace{1.0cm} & $0.95$ & $1.994$ & $2\times 10^{-3}$ &~~$6.83\times 10^{-5}$ &
$1.0\times 10^{-5}$\\

\multicolumn{1}{c}{\rule[-0.3cm]{0mm}{0.4cm}}&
\multicolumn{1}{c}{$0.85$}&
\multicolumn{1}{c}{$2.056$}&
\multicolumn{1}{c}{$2\times 10^{-3}$} &
\multicolumn{1}{c}{$~~2.17\times 10^{-4}$} &
\multicolumn{1}{c}{$1.5\times 10^{-5}$}\\ 
 
\hline

\multicolumn{1}{c}{\rule[0.1cm]{0mm}{0.5cm}}&
\multicolumn{1}{c}{$1.0$}&
\multicolumn{1}{c}{$2.001$}&
\multicolumn{1}{c}{$1\times 10^{-3}$} &
\multicolumn{1}{c}{$~~1.35\times 10^{-5}$} &
\multicolumn{1}{c}{$2.1\times 10^{-6}$}\\ 

$a/d$ = 1.0 & $0.98$ & $1.989$ & $1\times 10^{-3}$ & $~~4.84\times 10^{-5}$ &
$3.1\times 10^{-6}$\\
\hspace{1.0cm} & $0.95$ & $2.046$ & $1\times 10^{-3}$ & $-2.50\times 10^{-4}$ &
$8.6\times 10^{-6}$\\

\multicolumn{1}{c}{\rule[-0.3cm]{0mm}{0.4cm}}&
\multicolumn{1}{c}{$0.85$}&
\multicolumn{1}{c}{$2.064$}&
\multicolumn{1}{c}{$2\times 10^{-3}$}&
\multicolumn{1}{c}{$-4.78\times 10^{-5}$}&
\multicolumn{1}{c}{$1.7\times 10^{-5}$}\\ 

\hline

\multicolumn{1}{c}{\rule[0.1cm]{0mm}{0.5cm}}&
\multicolumn{1}{c}{$1.0$}&
\multicolumn{1}{c}{$1.987$}&
\multicolumn{1}{c}{$1\times 10^{-3}$} & 
\multicolumn{1}{c}{$~~3.22\times 10^{-5}$} &
\multicolumn{1}{c}{$2.0\times 10^{-6}$}\\

$a/d$ = 1.5 & $0.98$ & $2.025$ & $1\times 10^{-3}$ & $-5.38\times 10^{-5}$ &
$3.3\times 10^{-6}$\\
\hspace{1.0cm} & $0.95$ & $2.036$ & $2\times 10^{-3}$ & $-1.13\times 10^{-4}$ &
$9.2\times 10^{-6}$\\

\multicolumn{1}{c}{\rule[-0.1cm]{0mm}{0.35cm}}&
\multicolumn{1}{c}{$0.85$}&
\multicolumn{1}{c}{$2.075$}&
\multicolumn{1}{c}{$2\times 10^{-3}$}&
\multicolumn{1}{c}{$~~2.06\times 10^{-4}$}&
\multicolumn{1}{c}{$1.6\times 10^{-5}$}\\

\end{tabular}	
\end{ruledtabular}
\end{table}

\end{document}